# POSITION ANGLES AND ALIGNMENTS OF CLUSTERS OF GALAXIES

Manolis PLIONIS

*SISSA - International School for Advanced Studies,*
*Via Beirut 2-4, 34013 Trieste, Italy and*
*ICTP - International Centre for Theoretical Physics*

## Abstract

The position angles of a large number of Abell and Shectman clusters, identified in the Lick map as surface galaxy-density enhancements, are estimated. In total I determine the major axis orientation of 637 clusters out of which 448 are Shectman clusters (202 of which are also Abell clusters) and 189 are Abell clusters not originally detected by Shectman due to his adopted density threshold. Using published redshifts for 277 of these clusters I have detected strong nearest neighbour alignments over scales up to $\sim 15\ h^{-1}$ Mpc at a $\gtrsim 2.5 - 3\sigma$ significance level, while quite weak alignments are detected even up to $\sim 60\ h^{-1}$ Mpc. A more significant alignment signal ($\sim 4\sigma$) is detected among all neighbours residing in superclusters and having separations $\lesssim 10\ h^{-1}$ Mpc. Again, weaker but significant alignments are found when larger separations are considered. Since my cluster sample is neither volume limited nor redshift complete, a fact that would tend to wash-out any real alignment signal, the alignments detected should reflect a real and possibly a stronger underline effect.

# 1   Introduction

The issue of the elongation of clusters of galaxies (cf. Carter & Metcalfe 1980, Binggeli 1982, Plionis, Barrow & Frenk 1991) and their tendency to be aligned with their nearest neighbour and in many cases with the position angle of their first ranked galaxy is well studied and well documented, although conflicting results appear occasionally in the literature.

The alignment effect was first noted by Binggeli (1982) who claimed that clusters tend to be aligned with their neighbours over scales of 10-15 $h^{-1}$ Mpc. Since then, a number of authors have supported the reality of this effect (Rhee and Katgert 1987, Flin 1987, West 1989a,b, Lambas *et.al.* 1990, Rhee, van Haarlem & Katgert 1992), although doubts have been put forward about the significance and the strength of the effect (Struble & Peebles 1985; McMillan, Kowalski & Ulmer 1989; Fong, Stevenson & Shanks 1990). Note, however, that Fong *et.al.* searched for alignments between neighbouring clusters found in 2-dimensions. Any real signal could be diluted by projection and therefore one should be cautious on how to interpret these results. In fact, as it will be shown in what follows, a significant alignment signal between neighbours in 3 dimensions becomes insignificant, although still present, when the cluster pairs are chosen in angular space.

Further support for the reality of the alignment effect comes from the work of Argyres *et al* (1986) and Lambas, Groth & Peebles (1988) who found that the Lick galaxy counts around Abell clusters tend to be aligned with the cluster major axis out to $\sim 15$ $h^{-1}$ Mpc, especially for clusters in high density regions.

It was thought initially that the observed alignment effects would provide a very effective test to discriminate among different models of cosmic structure formation. In the pancake scenario (Zeldovich 1970, Doroshkevich, Shandarin & Saar 1978), like the Hot Dark Matter model, where clusters and galaxies form by fragmentation in already flattened sheet- and filament-like superclusters, one expects the clusters to be elongated and aligned. In accordance with this Dekel, West & Aarseth (1984) and West, Dekel & Oemler (1989) found that cluster alignments occur only when the initial fluctuation spectrum has a large coherence length as that expected in the HDM model. They where unable to reproduce the observed alignments in hierarchical clustering models, like the Cold Dark Matter (CDM) model, where the cosmic structures form by gravitational clustering from small to large scales (cf. Peebles 1982, Blumenthal *et.al.* 1984, Davis *et.al.* 1985, Frenk *et.al.* 1985, 1988). However, both the asphericity of clusters and the alignment effect could be produced in hierarchical models by a different mechanism. Tidal effects could influence the shapes of protoclusters and induce alignments. Indeed, Binney & Silk (1979) found that tidal effects between protostructures can induce flattening and prolate shapes for clusters with a mean ellipticity before virialization is $\langle \varepsilon \rangle \approx 0.5$, which is in good agreement with observations (Plionis, Barrow & Frenk 1991). However this issue is a controversial one since conflicting results have been presented in the literature. For example Barnes & Efstathiou (1987) using N-body simulations find that tidal interactions cannot induce alignments between neighboring protoclusters while Salvador-Sole & Solanes (1993) find, using analytical methods, that tides can induce the observed alignments.

Aside of these conflicting results and contrary to the numerical work of Dekel *et.al.* (1984) and West *et.al.* (1989), Bond (1987) has shown that within the framework of Gaussian statistics and if the clusters form at the peaks of the field then an alignment up to $\sim 20\ h^{-1}$ Mpc should be expected even in the CDM model. Soon after, analysis of new high-resolution CDM simulations showed that the alignment effect is present and even stronger than what was anticipated by the analytical work (West, Villumsen & Dekel 1991 and references therein). In view of these results, it would seem unavoidable to conclude that that cluster-cluster alignments is a generic feature of the cluster formation process which is independent of the specific model and form of the fluctuation spectrum. Therefore, cluster alignments cannot be used as an effective discriminant between models of structure formation. However, studying alignments and similar features of the distribution of matter on large scales, could provide interesting clues about the details of the cluster formation process and therefore it is essential:

1. To unambiguously determine whether alignments do occur in the real universe.

2. To find the amplitude of the effect and the scale over which it takes place.

3. To identify details of the alignment effect, for example whether galaxies within a cluster exhibit any alignment (cf. Struble 1990, van Kampen & Rhee 1990, Trevese, Cirimele & Flin 1992, Rhee, van Haarlem & Katgert 1992), which could then provide clues about the internal dynamics of the cluster (cf. Rhee & Roos 1990).

In this paper I present new cluster position angle determinations for a sample of 637 clusters of galaxies and I study the cluster alignment properties for a subsample of these clusters ($N = 277$) for which redshift information is available.

## 2   Cluster Position Angles

The cluster-finding algorithm, used to identify clusters from the Lick galaxy catalogue, and the resulting cluster catalogues were presented in Plionis, Barrow & Frenk (1991) [hereafter PBF]. The algorithm is based in identifying surface galaxy-density peaks above a given threshold and connecting in a unique cluster all neighbouring cells that also fulfill the overdensity criteria. A thorough statistical analysis of the distribution of these clusters was presented in Plionis & Borgani (1992), Borgani, Jing & Plionis (1992) and Borgani, Plionis & Valdarnini (1993). I therefore refer the interested reader to those articles as well as to Shane & Wirtanen (1967) and Seldner *et.al.* (1977) for issues related to the Lick galaxy catalogue and to Plionis (1988) for details of this particular use of the catalogue. I will just remind the reader that the Lick catalogue contains $\sim 810000$ unique galaxies with $m_b \lesssim 18.8$ which are binned in $10 \times 10$ arcmin$^2$ cells. The catalogue covers 70% of the sky and its characteristic depth is $210\ h^{-1}$ Mpc (Groth & Peebles 1977).

I determine the orientation of a cluster by estimating its position angle, $\theta$, measured relative to North in the anticlockwise direction. The principal axes of each cluster and their orientation are calculated by diagonalizing the $2 \times 2$ inertia tensor **I**:

$$I_{kl} = \sum_{i=1}^{n}(r_k r_l)_i m_i \qquad (1)$$

where $k, l = 1, 2$, $r_{1_i} = x_i$, $r_{2_i} = y_i$, $m_i$ is the galaxy number-count of the $i^{th}$ cell, $n$ is the number of $10' \times 10'$ cells for each cluster, and the origin of the coordinate axis is the centre of mass of the cluster. This method for determining cluster position angles is quite different from those used in most other studies (cf. Binggeli 1982) which measure the galaxy distribution within a fixed circular aperture around each cluster. This method has the disadvantage that it does not map the same area around each cluster but it has the advantage that it avoids any biases resulting from the use of a circular aperture (for a detailed discussion of this see Carter & Metcalfe 1980 and Binggeli 1982.)

My primary sample consists of all C36 clusters, ie., those with $\sigma/\langle\sigma\rangle = 3.6$ (see PBF), which sample a large enough area to make the determination of their position angle possible, ie., cover at least five $10 \times 10$ arcmin$^2$ cells. Note that the overdensity threshold used is the same as that used by Shectman (1985) and therefore these clusters correspond to Shectman clusters. In order to test the robustness of the position angle determination the clusters are traced to lower overdensity levels ($\sigma/\langle\sigma\rangle$=3, 2.5 and 1.8) and at each level their position angles are determined. Since, however, at lower overdensities there is a higher probability of the clusters being affected by projection effects that could distort their position angles and since projection effects should cause the apparent position of the cluster centre to vary from level to level, PBF attempted to minimize this problem adopting the following procedure. At each overdensity level the cluster centre of mass is estimated and only those levels are considered for which the cluster centre of mass, calculated at that overdensity, has shifted by < 10 arcmin from its original C36 position (ie., the cluster centre of mass should not move out of its original $10'$ cell). I also estimate the position angles of all Abell clusters, found at lower overdensity levels and which cover enough area to make the position angle determination possible. Note that in order to eliminate the gross effects of Galactic extinction the samples are limited to $|b| \geq 40°$.

In table 1 I present the C36 cluster position angles with their corresponding Shectman (1985) and Abell (1958) number. Where it was possible I also present the mean and standard deviation of their values from measurements at three, at least, overdensity levels. In such cases the median value is also listed. In table 1b I present the position angles of Abell clusters identified at $\sigma/\langle\sigma\rangle$=2.5 and determined at the same level, while in table 1c the listed position angles are of clusters identified at a level $\sigma/\langle\sigma\rangle$=1.8. Finally at table 1d I present position angles of clusters identified at the $\sigma/\langle\sigma\rangle$=2.5 but estimated at $\sigma/\langle\sigma\rangle$=1.8

## 3 Tests for systematic effects

PBF gave a thorough discussion of the most significant systematic biases that could affect the cluster shape parameters and found that the position angles remain mostly unaffected by the effect of the grid while when tracing the clusters to lower overdensity levels they found small variations of their position angle. However, when the cluster sampling area is small (ie, when few $10' \times 10'$ cells define the cluster), then due to the limiting geometry of the possible cell configuration, the position angles of such clusters tend to have a preferred range of values. Figure 1 shows the position angles as a function of the number of cells, defining the cluster sampling area, for clusters with only one determination of their position angle. Especially for $N_{cells} \leq 7$ the effect is evident. However, the effect is suppressed when more than one determinations of the position angle of a cluster is possible and in these cases the final value of $\theta$ is the average over all the determinations. This can be seen in figure 2 where we plot the position angles of clusters estimated in more than one overdensity level; panel (a) shows the values of $\langle\theta\rangle$ as a function of the minimum number of cells used while panel (b) the values of $\theta$ as a function of the maximum number of cells used for the $\langle\theta\rangle$ determination. I conclude that clusters with one position angle determination and $N_{cells} \leq 7$, have an uncertainty, due to the grid, of $\delta\theta \sim 30° - 35°$ and therefore the interested reader should be cautious of this effect and should consider their position angles rather as indicative.

To test whether this effect is significant and whether it could create a systematic orientation bias I calculate, following Struble & Peebles (1985), the following functions:

$$C_n = \sqrt{\frac{2}{N}} \sum_{i=1}^{N} \cos 2n\theta_i \qquad (2)$$

$$S_n = \sqrt{\frac{2}{N}} \sum_{i=1}^{N} \sin 2n\theta_i \qquad (3)$$

where $n \geq 1$ and $N$ is the total number of clusters in my sample. If the position angles are independently drawn from a uniform distribution between 0° and 180° then the $C_n$ and $S_n$ have zero mean and unit standard deviation. A systematic bias in the cluster orientations will manifest itself as a large value of $|C_n|$ and/or $|S_n|$. For all the 637 cluster position angles I obtain $|C_1|, |S_1| \lesssim 1.6$, $|C_2|$, $|S_2| \lesssim 1.2$ and $|C_3|, |S_3| \lesssim 1$ indicating that the distribution of position angles has no significant deviation from uniformity. However, when I use clusters with only one position angle determination and with $N_{cells} = 5$, I get $|S_2|, |C_2| \sim 2$ which reflects the grid effect mentioned before.

PBF found that when they compared the position angles of clusters in common with other studies, the typical uncertainty between the different measurements is $\delta\theta \sim 30° - 35°$, in agreement with West (1989). For example the position angle deviation for the clusters in common with Struble & Peebles (77 clusters) is $\delta\theta \sim 0° \pm 35.5°$ while when I compare with 61 common clusters of Rhee *et.al.* (1992) I find $\delta\theta \sim -1.4° \pm 40.5°$ for those determined by the Fourier method while I find slightly worse results for their other methods. Interestingly, I find the best agreement when I compare my cluster position angles with their $1^{st}$ rank galaxy position angles ($\delta\theta \sim 0.5° \pm 39°$).

Note, however, that PBF found that if the comparison is restricted only to those common clusters for which their estimate of the position angle is based on an area similar in size to the circular aperture used in other studies then the position angle deviation is reduced by $\sim 50\%$. This shows that the different position angle estimation methods lead to similar values when applied to similar regions around the clusters.

# 4  Alignments of Clusters

In order to search for alignments in my sample I collected all the published cluster redshifts to end up with 277 clusters in the North and South galactic hemisphere. For this sample I find $|C_1|, |S_1| \leq 1.2$, $|C_2|, |S_2| \leq 0.2$ and $|C_3|, |S_3| \leq 0.9$ which again shows that this sample is free of orientation bias. Note that my sample is not volume limited, a major drawback in such a study, nor is it redshift complete a fact that could hinder me from reaching any strong conclusion, especially if I would not detect any alignments.

The distance to each cluster is determined by the usual relation (Mattig 1958):

$$R = \frac{c}{H_\circ q_\circ^2 (1+z)^2}[q_\circ z + (q_\circ - 1)[(1 + 2q_\circ z)^{\frac{1}{2}} - 1]] \tag{4}$$

with $q_\circ = 0.2$. The cluster coordinate system is transformed to a Cartesian one using the transformations: $x = R \cos b \sin l$, $y = R \cos b \cos l$ and $z = R \sin b$. In this coordinate system the relative distance $D_{ij}$ of each cluster pair are found by $D_{ij} = (\delta x_{ij}^2 + \delta y_{ij}^2 + \delta z_{ij}^2)^{1/2}$. I determine the position angle $\phi_{ij}$ of each cluster, $i$, relative to the direction of a neighbour, $j$, as the acute angle between the major axis of the cluster and the great circle connecting the two clusters. Firstly the position angle of the cluster-pair separation, $\phi_{is}$, at the position of the primary cluster is found by using straightforward spherical trigonometry and finally $\phi_{ij} = \theta_i - \phi_{is}$, where $\theta_i$ is the position angle of the primary cluster.

The mean $\phi_{ij}$ in an isotropic distribution with large $N$ would be equal to 45°. A deviation from this value, in a bias free sample, would be an indication of an alignment/misalignment effect. A useful measure of such an effect is given by Struble & Peebles (1985):

$$\delta = \sum_{i=1}^{N} \frac{\phi_{ij}}{N} - 45 \tag{5}$$

If the values of $\phi_{ij}$ are isotropically distributed between 0° and 90° then for large $N$, $\langle \delta \rangle = 0$ and the standard deviation is:

$$\sigma = \frac{90}{\sqrt{12N}} \tag{6}$$

A negative value of $\delta$ would indicate alignment and a positive one a misalignment. We must consider though that a comparison of observations with these formulas should take place only if the determination of position angles does not entail systematic errors. If there is a systematic bias of the cluster position angles towards a preferred direction it could produce a false alignment or misalignment signal. This, however, is not the case for my samples since it was found, in section 3, that such a bias is not present. I searched for two types of cluster alignments:

1. Nearest-Neighbour alignments (N-N)

2. All neighbour alignments within superclusters (A-N)

## 4.1 Nearest-Neighbour Alignments

To study this type of alignments I found for each cluster $i$ a neighbour $j$ for which their distance is $\min[D_{ij}]$. For such pairs we then calculate $\phi_{ij} (\equiv \phi_{nn})$ and estimate the alignment signal $\delta$ and its deviation $\sigma$. Note that $\mu \equiv \langle \min[D_{ij}] \rangle$ is $\sim 25\ h^{-1}$ Mpc, a rather large value but consider that the sample is not volume limited and there are a few clusters with very large redshifts. If we limit the N-N separations to a maximum value of 30 or 50 $h^{-1}$ Mpc then $\mu \sim 14$ and 20 $h^{-1}$ Mpc, respectively.

In Table 2 we present our results for different limiting cluster pair separations, $D_{lim}$. A quite significant signal (2.5$\sigma$ effect) is found when $D_{lim} = 15\ h^{-1}$ Mpc which persists, although having a smaller value of $\delta$, even when very large separations are considered ($D_{lim} = 50$ - $60\ h^{-1}$ Mpc). Figure 3 shows the frequency distribution of the $\phi_{nn}$ angles. Figure 3a shows the results for $D_{lim} = 10$ and 15 $h^{-1}$ Mpc (solid and dashed lines respectively) while figure 3b shows the results when all the N-N separations are considered. A clear excess of small $\phi_{nn}$ values is present in all cases, being more pronounced at smaller values of $D_{lim}$.

I attempted to further test the significance of my result by testing whether the observed $\phi_{nn}$ distribution could have been drawn from a uniform one. To this end we perform a Kolmogorov-Smirnov two-sided test and we find that the probability of the observed distribution being drawn from a uniform parent distribution is $\sim 4 \times 10^{-2}$, $\sim 3 \times 10^{-2}$ and $\sim 5 \times 10^{-2}$ for $D_{lim} = 10$, 15 and 60 $h^{-1}$ Mpc respectively.

## 4.2 All-Neighbour Alignments within Superclusters

I have used the above procedure to study alignments between all possible cluster-pairs that lie within the same supercluster. Binggeli (1982) has claimed a positive alignment signal of this type, although at a low significance level. Similarly, West (1989b) has found that such alignments occur over scales of *at least* 30 $h^{-1}$ Mpc and maybe over even larger distances, while Plionis, Valdarnini & Jing (1992), using Abell clusters and a subset of the position angles presented here found that indeed there is a strong alignment signal between all neighbours within superclusters which seems also to be correlated with the shape of the supercluster, being stronger in prolate configurations.

I define superclusters using a friend of friend algorithm and two percolation radii, $R_p$, of 35 $h^{-1}$ and 25 $h^{-1}$ Mpc length. Since my main results seem to be insensitive to these two choices of $R_p$, I will present only those for the $R_p = 35\ h^{-1}$ case. For this value of $R_p$ I obtain 33 superclusters (with more than two members) containing 209 out of the 277 clusters. The alignment signal (eq. 5) is used but note that I used clusters that belong into superclusters of any membership number above unity, while Plionis *et.al.* (1992) used supercluster with more than 8 members (because they wanted to determine also their shapes).

In table 3 I present the values of the alignment signal $\delta$ and its significance for different cluster separation ranges. Strong alignments occur in my sample only for $D_{lim} = 10\ h^{-1}$ Mpc while weaker but still significant alignments are present even when $D_{lim} \sim 150\ h^{-1}$ Mpc. Remember, however, that my samples are not volume limited nor redshift complete and therefore even such an alignment signal should be considered as an indication of a probably larger and more significant alignment effect. In figure 4 I present the frequency distribution of the $\phi_{ij}$ angles at four cluster separation ranges. The strong alignment effect at the $D \leq 10\ h^{-1}$ Mpc range is evident. Our results are in agreement with those of West (1989b) and Plionis *et.al.* (1992) even more so if we take into account that our samples are not as complete, a fact that would tend wash-out any inherent alignment signal.

## 4.3 Possible systematic biases

The observed alignments could be artificial, in principle, resulting from the fact that clusters which are near in space are in many occasions also near on the sky and therefore their galaxy density envelopes could overlap in the angular projection. This could produce a directional bias when determining their position angles and thus a preferred orientation along the direction of a neighbour, in particular of the nearest neighbour. A manifestation of such an effect would be to find a stronger and more significant alignment signal between nearest neighbours in angular space rather than in real 3-D space. Since however, a number of clusters that are neighbours in real space are neighbours also in angular space some weak alignment signal should survive the projection. To test whether this effect is responsible for the alignment signal found, I searched for nearest neighbour alignments in my sample but using angular separations instead of spatial ones. I found for all separations a significantly weaker alignment signal. For example when the angular separations are $\leq 1.4°$ (for which I have in total 66 separation, a value similar to that for $D \leq 10\ h^{-1}$ Mpc) I obtain a weaker and less significant alignment signal ($\delta = -4.1 \pm 3.2$) with a 0.15 probability of the distribution of $\phi_{nn}$ being drawn from a uniform one, confirming that the observed alignment signal in 3-D is not caused by the above mentioned bias.

Furthermore, to test whether the grid effect, mentioned in section 3, has induced the detected cluster alignments, I repeated the N-N analysis but after having reshuffled the cluster position angles. I performed a 100 such reshufflings and I find, for all separations considered, $\langle \delta \rangle = 0$ with a standard deviation equal to that obtained from eq. 6, which indicates that the grid effect is not responsible for the detected cluster alignment signal.

## 5 Conclusions

I have presented a large number of new cluster position angles. In total 637 position angles were estimated. Using available redshifts for a subset of these clusters I found a strong alignment signal ($\sim 2.5 - 3\sigma$) between nearest cluster neighbours when their separations are $\leq 15\ h^{-1}$ Mpc, as well as a stronger ($\sim 4\sigma$) signal between all neighbours residing in superclusters when separations $\leq 10\ h^{-1}$ Mpc are considered. A quite weak but significant alignment signal persists even when considering larger separations (up to 150 $h^{-1}$ Mpc).

**Table 1a**: Mean position angles (column 5) of Abell and Shectman clusters (identified at the 3.6 level). We present the number of overdensity levels used to estimate the position angles (column 3), the minimum and maximum number of $10' \times 10'$ cells, $N_{cells}$, defining the cluster sampling area [at the different overdensity levels] (column 4), while for the cases were the position angle ($\theta$) was determined in 3 or more overdensity levels we list also $\sigma_\theta$ (column 6) and the median value (column 7) [in those cases where 4 levels were used we averge the two central $\theta$ values].

| Shectman | Abell | No of levels | min-max $N_{cells}$ | $\theta$ | $\sigma_\theta$ | Med[$\theta$] |
|---|---|---|---|---|---|---|
| 1  | 724  | 1 | 6     | 177.2 | ... | ... |
| 2  | 727  | 2 | 10-15 | 65.8  | ... | ... |
| 4  | -    | 2 | 7-12  | 80.4  | ... | ... |
| 5  | 757  | 2 | 5-9   | 79.3  | ... | ... |
| 6  | -    | 2 | 6-14  | 36.5  | ... | ... |
| 8  | 819  | 1 | 7     | 111.0 | ... | ... |
| 9  | 858  | 1 | 7     | 96.1  | ... | ... |
| 10 | -    | 2 | 7-10  | 103.1 | ... | ... |
| 11 | 879  | 1 | 8     | 132.4 | ... | ... |
| 12 | -    | 1 | 6     | 166.1 | ... | ... |
| 16 | 912  | 1 | 6     | 0.6   | ... | ... |
| 18 | 921  | 2 | 5-9   | 118.1 | ... | ... |
| 21 | 949  | 2 | 7-9   | 40.5  | ... | ... |
| 22 | -    | 3 | 8-20  | 165.9 | 14.6 | 157.5 |
| 23 | 952  | 2 | 7-10  | 119.8 | ... | ... |
| 24 | -    | 3 | 5-13  | 115.4 | 5.7  | 114.4 |
| 25 | -    | 3 | 6-13  | 19.4  | 19.2 | 20.1 |
| 26 | 978  | 3 | 8-13  | 169.3 | 3.5  | 167.7 |
| 27 | -    | 2 | 5-7   | 47.0  | ... | ... |
| 28 | 986  | 2 | 5-8   | 119.9 | ... | ... |
| 30 | -    | 2 | 7-11  | 97.8  | ... | ... |
| 31 | -    | 2 | 8-15  | 157.2 | ... | ... |
| 33 | 993  | 3 | 9-27  | 114.9 | 3.4  | 113.8 |
| 35 | -    | 3 | 6-19  | 133.7 | 7.1  | 131.8 |
| 37 | -    | 2 | 6-14  | 22.0  | ... | ... |
| 38 | -    | 1 | 6     | 109.0 | ... | ... |
| 39 | 1020 | 4 | 10-40 | 137.1 | 19.2 | 144.6 |
| 42 | -    | 2 | 5-28  | 102.5 | ... | ... |
| 43 | -    | 1 | 7     | 60.9  | ... | ... |
| 45 | -    | 1 | 6     | 44.5  | ... | ... |
| 46 | 1033 | 1 | 8     | 43.9  | ... | ... |
| 47 | 1035 | 4 | 5-84  | 93.8  | 21.1 | 90.6 |
| 48 | -    | 2 | 5-10  | 131.6 | ... | ... |
| 50 | 1050 | 3 | 5-13  | 76.6  | 6.5  | 74.2 |

| Shectman | Abell | No of levels | min-max $N_{cells}$ | $\theta$ | $\sigma_\theta$ | Med[$\theta$] |
|---|---|---|---|---|---|---|
| 52 | 1066 | 4 | 6-27 | 9.6 | 24.0 | 6.7 |
| 53 | 1067 | 1 | 8 | 45.0 | ... | ... |
| 54 | - | 1 | 10 | 157.4 | ... | ... |
| 55 | - | 2 | 6-10 | 116.9 | ... | ... |
| 57 | 1085 | 1 | 8 | 176.6 | ... | ... |
| 59 | 1100 | 2 | 5-10 | 101.4 | ... | ... |
| 60 | - | 2 | 13-25 | 155.2 | ... | ... |
| 61 | - | 3 | 5-11 | 102.2 | 6.4 | 103.6 |
| 62 | - | 2 | 7-9 | 119.7 | ... | ... |
| 63 | 1126 | 1 | 5 | 161.8 | ... | ... |
| 65 | - | 2 | 6-11 | 91.5 | ... | ... |
| 66 | 1139 | 2 | 6-14 | 112.5 | ... | ... |
| 67 | - | 1 | 10 | 44.5 | ... | ... |
| 68 | 1149 | 2 | 7-9 | 138.2 | ... | ... |
| 69 | - | 2 | 6-9 | 61.5 | ... | ... |
| 70 | - | 2 | 5-11 | 135.6 | ... | ... |
| 71 | 1169 | 4 | 9-89 | 52.1 | 33.8 | 57.5 |
| 72 | 1168 | 2 | 5-11 | 44.0 | ... | ... |
| 73 | 1173 | 2 | 5-9 | 102.5 | ... | ... |
| 74 | 1177 | 2 | 8-16 | 13.9 | ... | ... |
| 75 | 1185 | 3 | 14-51 | 169.4 | 47.4 | 145.4 |
| 76 | - | 1 | 13 | 168.5 | ... | ... |
| 77 | 1190 | 1 | 5 | 180.0 | ... | ... |
| 78 | 1187 | 1 | 11 | 156.7 | ... | ... |
| 82 | 1205 | 3 | 8-16 | 89.3 | 7.4 | 91.4 |
| 83 | - | 1 | 7 | 61.5 | ... | ... |
| 84 | - | 1 | 9 | 21.3 | ... | ... |
| 85 | - | 2 | 11-19 | 54.3 | ... | ... |
| 86 | 1213 | 3 | 12-29 | 96.0 | 10.4 | 90.2 |
| 87 | - | 1 | 5 | 57.6 | ... | ... |
| 88 | - | 3 | 5-16 | 116.0 | 49.4 | 107.1 |
| 89 | - | 4 | 8-26 | 50.8 | 29.0 | 43.2 |
| 90 | - | 3 | 8-20 | 6.4 | 53.5 | 2.5 |
| 91 | 1235 | 2 | 10-11 | 73.6 | ... | ... |

| Shectman | Abell | No of levels | min-max $N_{cells}$ | $\theta$ | $\sigma_\theta$ | Med[$\theta$] |
|---|---|---|---|---|---|---|
| 97  | -    | 1 | 6     | 30.8  | ...  | ...   |
| 99  | -    | 1 | 8     | 135.0 | ...  | ...   |
| 100 | 1267 | 2 | 6-9   | 93.3  | ...  | ...   |
| 102 | -    | 1 | 7     | 7.7   | ...  | ...   |
| 103 | -    | 2 | 8-13  | 33.7  | ...  | ...   |
| 104 | -    | 1 | 15    | 107.6 | ...  | ...   |
| 105 | 1291 | 3 | 7-15  | 172.6 | 21.5 | 1.6   |
| 106 | -    | 2 | 5-7   | 103.2 | ...  | ...   |
| 107 | 1307 | 3 | 5-12  | 6.9   | 9.3  | 5.0   |
| 108 | 1308 | 1 | 7     | 80.2  | ...  | ...   |
| 109 | -    | 1 | 7     | 135.8 | ...  | ...   |
| 110 | 1314 | 4 | 10-26 | 75.0  | 2.0  | 74.4  |
| 111 | -    | 2 | 8-18  | 97.2  | ...  | ...   |
| 112 | 1317 | 3 | 6-13  | 42.6  | 38.2 | 21.2  |
| 113 | -    | 2 | 7-13  | 85.2  | ...  | ...   |
| 114 | 1318 | 1 | 8     | 1.6   | ...  | ...   |
| 115 | -    | 1 | 5     | 105.5 | ...  | ...   |
| 117 | -    | 1 | 5     | 18.0  | ...  | ...   |
| 118 | 1332 | 1 | 7     | 135.8 | ...  | ...   |
| 119 | 1336 | 2 | 8-11  | 120.4 | ...  | ...   |
| 121 | 1341 | 4 | 13-45 | 159.6 | 20.9 | 160.1 |
| 122 | 1346 | 2 | 17-26 | 164.8 | ...  | ...   |
| 124 | 1364 | 1 | 12    | 120.9 | ...  | ...   |
| 125 | 1362 | 3 | 9-26  | 149.8 | 5.7  | 148.2 |
| 126 | -    | 1 | 7     | 78.3  | ...  | ...   |
| 131 | 1367 | 3 | 21-46 | 156.8 | 20.9 | 150.6 |
| 128 | -    | 1 | 6     | 90.0  | ...  | ...   |
| 129 | -    | 2 | 9-14  | 118.2 | ...  | ...   |
| 130 | 1365 | 3 | 8-12  | 11.7  | 12.4 | 7.3   |
| 134 | -    | 1 | 5     | 20.4  | ...  | ...   |
| 135 | 1371 | 2 | 8-17  | 100.1 | ...  | ...   |
| 136 | -    | 3 | 9-20  | 126.5 | 1.4  | 126.4 |
| 137 | 1377 | 3 | 10-19 | 165.3 | 13.4 | 165.5 |
| 138 | 1380 | 1 | 7     | 173.9 | ...  | ...   |

| Shectman | Abell | No of levels | min-max $N_{cells}$ | $\theta$ | $\sigma_\theta$ | Med[$\theta$] |
|---|---|---|---|---|---|---|
| 139 | - | 1 | 5 | 17.4 | ... | ... |
| 140 | - | 1 | 5 | 72.2 | ... | ... |
| 142 | 1383 | 2 | 5-8 | 78.9 | ... | ... |
| 144 | - | 2 | 8-12 | 148.8 | ... | ... |
| 145 | - | 1 | 6 | 90.0 | ... | ... |
| 148 | - | 3 | 5-17 | 152.1 | 11.8 | 158.4 |
| 151 | 1399 | 1 | 9 | 149.4 | ... | ... |
| 153 | - | 1 | 6 | 0.8 | ... | ... |
| 154 | - | 2 | 5-9 | 92.8 | ... | ... |
| 155 | - | 2 | 7-17 | 135.0 | ... | ... |
| 156 | - | 1 | 9 | 117.1 | ... | ... |
| 157 | 1407 | 1 | 6 | 2.2 | ... | ... |
| 158 | 1413 | 2 | 9-23 | 107.0 | ... | ... |
| 159 | 1416 | 1 | 5 | 73.9 | ... | ... |
| 161 | - | 3 | 5-13 | 89.3 | 13.1 | 95.8 |
| 162 | 1424 | 3 | 5-12 | 57.0 | 9.8 | 58.6 |
| 163 | 1436 | 3 | 10-14 | 29.5 | 3.3 | 31.4 |
| 164 | 1448 | 2 | 5-6 | 121.3 | ... | ... |
| 165 | 1452 | 1 | 5 | 109.4 | ... | ... |
| 167 | - | 2 | 8-11 | 106.7 | ... | ... |
| 169 | - | 2 | 5-7 | 165.8 | ... | ... |
| 170 | 1468 | 2 | 5-7 | 37.4 | ... | ... |
| 171 | - | 3 | 9-32 | 4.7 | 7.4 | 8.1 |
| 174 | 1502 | 2 | 7-11 | 7.1 | ... | ... |
| 175 | - | 3 | 10-24 | 143.6 | 15.8 | 138.0 |
| 176 | 1507 | 3 | 7-10 | 60.7 | 11.2 | 64.8 |
| 177 | - | 1 | 7 | 60.3 | ... | ... |
| 178 | - | 3 | 8-18 | 16.3 | 22.3 | 13.7 |
| 180 | 1516 | 2 | 9-20 | 89.6 | ... | ... |
| 181 | 1517 | 2 | 9-20 | 158.0 | ... | ... |
| 183 | 1520 | 4 | 17-52 | 8.6 | 19.0 | 0.5 |
| 184 | - | 1 | 5 | 19.5 | ... | ... |
| 185 | - | 3 | 5-13 | 173.6 | 45.7 | 162.8 |
| 186 | 1535 | 3 | 7-13 | 141.2 | 41.2 | 129.9 |

| Shectman | Abell | No of levels | min-max $N_{cells}$ | $\theta$ | $\sigma_\theta$ | Med[$\theta$] |
|---|---|---|---|---|---|---|
| 187 | 1541 | 4 | 8-44 | 117.0 | 9.3 | 114.2 |
| 189 | - | 2 | 6-14 | 160.4 | ... | ... |
| 190 | 1552 | 4 | 5-40 | 163.3 | 42.6 | 165.0 |
| 192 | 1553 | 1 | 8 | 137.1 | ... | ... |
| 193 | 1555 | 3 | 6-20 | 96.0 | 7.9 | 94.7 |
| 194 | - | 2 | 5-12 | 113.3 | ... | ... |
| 195 | 1564 | 2 | 5-10 | 111.1 | ... | ... |
| 196 | 1569 | 3 | 7-18 | 104.1 | 4.7 | 104.6 |
| 197 | - | 1 | 5 | 162.0 | ... | ... |
| 198 | - | 1 | 15 | 148.6 | ... | ... |
| 199 | 1589 | 3 | 7-15 | 142.0 | 34.5 | 152.9 |
| 201 | - | 2 | 7-22 | 142.9 | ... | ... |
| 203 | - | 1 | 8 | 177.8 | ... | ... |
| 204 | 1620 | 3 | 8-23 | 80.5 | 10.0 | 78.9 |
| 208 | - | 2 | 7-14 | 48.5 | ... | ... |
| 209 | 1631 | 3 | 29-73 | 142.5 | 5.7 | 142.6 |
| 210 | - | 2 | 5-8 | 120.6 | ... | ... |
| 211 | - | 1 | 6 | 64.1 | ... | ... |
| 212 | 1638 | 3 | 5-14 | 26.9 | 5.1 | 29.4 |
| 213 | - | 3 | 5-10 | 91.3 | 3.8 | 90.8 |
| 215 | 1644 | 3 | 20-48 | 86.4 | 42.8 | 91.8 |
| 216 | - | 3 | 21-45 | 128.8 | 1.2 | 128.1 |
| 218 | - | 1 | 9 | 105.7 | ... | ... |
| 219 | 1650 | 4 | 6-17 | 145.6 | 6.2 | 144.6 |
| 221 | 1651 | 2 | 7-13 | 59.3 | ... | ... |
| 222 | 1656 | 3 | 32-67 | 71.5 | 1.0 | 71.8 |
| 224 | 1663 | 3 | 7-14 | 30.7 | 20.0 | 39.0 |
| 226 | - | 2 | 6-12 | 73.8 | ... | ... |
| 227 | - | 3 | 6-46 | 91.7 | 41.0 | 109.1 |
| 228 | - | 2 | 5-13 | 113.1 | ... | ... |
| 229 | 1668 | 1 | 6 | 48.4 | ... | ... |
| 230 | - | 2 | 5-7 | 5.8 | ... | ... |
| 231 | - | 2 | 9-36 | 101.1 | ... | ... |
| 232 | - | 2 | 6-12 | 37.6 | ... | ... |

| Shectman | Abell | No of levels | min-max $N_{cells}$ | $\theta$ | $\sigma_\theta$ | Med[$\theta$] |
|---|---|---|---|---|---|---|
| 234 | 1691 | 3 | 14-26 | 121.2 | 0.7 | 120.8 |
| 235 | - | 3 | 6-14 | 130.9 | 23.9 | 138.0 |
| 236 | 1706 | 2 | 10-16 | 121.8 | ... | ... |
| 237 | - | 1 | 6 | 45.5 | ... | ... |
| 238 | 1709 | 2 | 5-8 | 158.0 | ... | ... |
| 240 | 1711 | 1 | 5 | 72.2 | ... | ... |
| 241 | - | 4 | 6-75 | 127.2 | 22.0 | 125.9 |
| 243 | - | 1 | 9 | 158.9 | ... | ... |
| 244 | - | 3 | 6-16 | 106.9 | 25.7 | 104.8 |
| 246 | - | 2 | 6-12 | 22.5 | ... | ... |
| 247 | - | 3 | 5-27 | 69.2 | 18.9 | 71.4 |
| 248 | 1738 | 3 | 5-17 | 113.0 | 16.0 | 108.2 |
| 249 | 1749 | 4 | 8-19 | 125.1 | 9.6 | 122.4 |
| 250 | - | 3 | 10-27 | 168.2 | 13.6 | 169.4 |
| 253 | - | 4 | 7-19 | 79.7 | 22.2 | 90.0 |
| 254 | - | 2 | 5-8 | 35.5 | ... | ... |
| 256 | - | 1 | 8 | 50.0 | ... | ... |
| 257 | 1764 | 1 | 5 | 161.5 | ... | ... |
| 258 | - | 1 | 9 | 20.8 | ... | ... |
| 259 | 1767 | 1 | 8 | 33.4 | ... | ... |
| 261 | - | 2 | 6-8 | 132.8 | ... | ... |
| 262 | - | 1 | 7 | 59.0 | ... | ... |
| 263 | 1775 | 3 | 12-41 | 88.3 | 6.0 | 90.9 |
| 264 | 1773 | 3 | 5-14 | 61.1 | 28.4 | 71.1 |
| 265 | - | 3 | 11-44 | 89.0 | 21.5 | 97.5 |
| 266 | 1778 | 4 | 6-37 | 86.8 | 12.1 | 84.2 |
| 267 | 1783 | 4 | 6-18 | 77.0 | 36.3 | 65.3 |
| 268 | 1780 | 1 | 5 | 109.2 | ... | ... |
| 270 | 1793 | 1 | 6 | 92.6 | ... | ... |
| 271 | - | 2 | 8-10 | 124.7 | ... | ... |
| 272 | 1795 | 3 | 11-21 | 13.8 | 12.6 | 16.5 |
| 273 | 1800 | 3 | 9-18 | 48.9 | 22.3 | 57.1 |
| 274 | - | 2 | 6-12 | 62.2 | ... | ... |
| 275 | 1797 | 2 | 9-16 | 135.4 | ... | ... |

| Shectman | Abell | No of levels | min-max $N_{cells}$ | $\theta$ | $\sigma_\theta$ | Med$[\theta]$ |
|---|---|---|---|---|---|---|
| 276 | - | 1 | 6 | 179.2 | ... | ... |
| 277 | - | 3 | 6-25 | 26.9 | 8.4 | 23.5 |
| 278 | - | 2 | 6-9 | 94.8 | ... | ... |
| 279 | - | 3 | 5-14 | 85.9 | 13.4 | 93.0 |
| 280 | 1809 | 4 | 8-12 | 83.8 | 6.5 | 86.8 |
| 282 | - | 4 | 5-37 | 22.0 | 10.8 | 21.1 |
| 283 | 1812 | 3 | 5-11 | 14.7 | 5.4 | 16.6 |
| 285 | - | 3 | 6-28 | 164.9 | 12.6 | 162.2 |
| 286 | - | 1 | 5 | 158.9 | ... | ... |
| 287 | - | 2 | 6-10 | 81.4 | ... | ... |
| 288 | 1825 | 1 | 5 | 161.8 | ... | ... |
| 289 | 1834 | 3 | 5-14 | 157.2 | 18.1 | 161.6 |
| 290 | 1831 | 4 | 11-74 | 130.9 | 35.4 | 124.9 |
| 291 | 1836 | 3 | 10-25 | 18.3 | 15.9 | 25.0 |
| 293 | 1852 | 3 | 6-15 | 1.2 | 10.1 | ... |
| 294 | - | 4 | 6-35 | 88.9 | 9.6 | 88.0 |
| 296 | - | 1 | 8 | 70.7 | ... | ... |
| 297 | - | 1 | 6 | 4.3 | ... | ... |
| 298 | - | 1 | 6 | 91.6 | ... | ... |
| 300 | 1873 | 1 | 8 | 178.5 | ... | ... |
| 301 | - | 2 | 5-27 | 171.9 | ... | ... |
| 304 | - | 2 | 5-6 | 29.4 | ... | ... |
| 305 | - | 1 | 6 | 91.5 | ... | ... |
| 306 | 1882 | 2 | 8-16 | 90.7 | ... | ... |
| 307 | - | 2 | 7-14 | 71.3 | ... | ... |
| 308 | 1890 | 3 | 8-11 | 26.2 | 34.9 | 46.3 |
| 309 | - | 3 | 11-18 | 11.9 | 7.2 | 14.8 |
| 310 | - | 2 | 5-12 | 32.3 | ... | ... |
| 312 | - | 3 | 6-12 | 61.1 | 15.9 | 53.0 |
| 313 | - | 2 | 6-11 | 31.9 | ... | ... |
| 314 | - | 3 | 9-14 | 111.2 | 4.4 | 110.0 |
| 315 | 1899 | 1 | 6 | 91.3 | ... | ... |
| 316 | 1904 | 4 | 7-20 | 25.2 | 14.8 | 23.7 |
| 317 | 1906 | 2 | 5-8 | 63.3 | ... | ... |

| Shectman | Abell | No of levels | min-max $N_{cells}$ | $\theta$ | $\sigma_\theta$ | Med[$\theta$] |
|---|---|---|---|---|---|---|
| 318 | 1908 | 2 | 11-18 | 20.7 | ... | ... |
| 319 | - | 1 | 6 | 88.4 | ... | ... |
| 321 | 1913 | 4 | 14-91 | 141.9 | 26.9 | 132.8 |
| 322 | - | 1 | 5 | 69.7 | ... | ... |
| 323 | - | 2 | 5-7 | 10.0 | ... | ... |
| 324 | - | 1 | 12 | 101.0 | ... | ... |
| 327 | - | 3 | 5-20 | 99.8 | 8.5 | 96.9 |
| 329 | - | 1 | 6 | 59.0 | ... | ... |
| 331 | - | 3 | 5-15 | 21.2 | 6.1 | 18.5 |
| 332 | - | 3 | 5-16 | 58.3 | 42.5 | 70.6 |
| 333 | 1964 | 4 | 5-13 | 53.3 | 13.4 | 49.3 |
| 334 | - | 1 | 9 | 108.6 | ... | ... |
| 335 | - | 1 | 7 | 81.7 | ... | ... |
| 339 | 1983 | 3 | 11-24 | 157.4 | 7.1 | 154.6 |
| 340 | 1986 | 1 | 5 | 20.8 | ... | ... |
| 341 | 1991 | 4 | 9-44 | 5.0 | 25.8 | 4.5 |
| 342 | - | 2 | 6-9 | 50.5 | ... | ... |
| 343 | - | 2 | 5-8 | 120.3 | ... | ... |
| 345 | 2020 | 3 | 5-13 | 17.6 | 8.7 | 18.8 |
| 346 | - | 1 | 5 | 161.9 | ... | ... |
| 347 | 2022 | 4 | 7-34 | 112.3 | 22.1 | 112.1 |
| 348 | - | 2 | 8-10 | 53.3 | ... | ... |
| 350 | - | 1 | 5 | 20.1 | ... | ... |
| 351 | - | 1 | 15 | 136.5 | ... | ... |
| 352 | - | 2 | 7-14 | 19.1 | ... | ... |
| 353 | 2028 | 3 | 5-37 | 93.4 | 19.4 | 103.7 |
| 354 | 2029 | 1 | 23 | 5.1 | ... | ... |
| 362 | - | 3 | 10-19 | 124.6 | 3.7 | 126.4 |
| 365 | 2048 | 1 | 45 | 106.2 | ... | ... |
| 366 | 2052 | 2 | 8-12 | 46.4 | ... | ... |
| 368 | - | 1 | 9 | 3.0 | ... | ... |
| 369 | 2055 | 1 | 6 | 139.4 | ... | ... |
| 370 | - | 3 | 5-34 | 89.9 | 21.6 | 83.8 |
| 371 | - | 1 | 31 | 138.4 | ... | ... |

| Shectman | Abell | No of levels | min-max $N_{cells}$ | $\theta$ | $\sigma_\theta$ | Med[$\theta$] |
|---|---|---|---|---|---|---|
| 372 | - | 1 | 6 | 137.1 | ... | ... |
| 374 | - | 1 | 8 | 135.5 | ... | ... |
| 375 | - | 1 | 8 | 90.5 | ... | ... |
| 376 | 2062 | 3 | 5-8 | 83.9 | 44.0 | 72.1 |
| 377 | 2061 | 3 | 20-39 | 35.2 | 6.7 | 32.8 |
| 378 | 2063 | 1 | 9 | 45.1 | ... | ... |
| 380 | 2065 | 2 | 25-39 | 176.1 | ... | ... |
| 383 | 2069 | 1 | 16 | 116.4 | ... | ... |
| 384 | - | 1 | 6 | 89.5 | ... | ... |
| 385 | 2079 | 1 | 10 | 37.0 | ... | ... |
| 386 | 2083 | 4 | 6-20 | 39.9 | 2.8 | 39.4 |
| 389 | - | 3 | 6-22 | 17.9 | 15.3 | 23.4 |
| 390 | 2089 | 1 | 6 | 87.8 | ... | ... |
| 391 | 2092 | 3 | 8-29 | 24.9 | 18.8 | 32.9 |
| 392 | 2107 | 3 | 7-14 | 52.4 | 24.0 | 46.5 |
| 393 | 2122 | 4 | 5-20 | 150.9 | 16.1 | 156.2 |
| 395 | - | 2 | 6-7 | 124.7 | ... | ... |
| 396 | - | 2 | 5-10 | 177.7 | ... | ... |
| 397 | - | 3 | 6-16 | 90.8 | 2.9 | 89.4 |
| 398 | - | 1 | 11 | 13.8 | ... | ... |
| 400 | 2142 | 2 | 14-25 | 145.1 | ... | ... |
| 401 | - | 2 | 7-17 | 139.1 | ... | ... |
| 403 | 2149 | 2 | 5-8 | 111.4 | ... | ... |
| 404 | - | 3 | 46-74 | 126.7 | 25.1 | 112.9 |
| 407 | 2151 | 3 | 21-34 | 169.4 | 29.8 | 3.6 |
| 409 | 2162 | 3 | 8-17 | 1.7 | 10.2 | 7.3 |
| 411 | - | 4 | 7-23 | 82.2 | 8.3 | 83.3 |
| 414 | - | 3 | 7-22 | 54.1 | 46.5 | 72.0 |
| 415 | - | 1 | 8 | 124.6 | ... | ... |
| 416 | 2199 | 1 | 17 | 47.2 | ... | ... |
| 417 | - | 1 | 8 | 124.6 | ... | ... |
| 418 | - | 3 | 5-14 | 68.8 | 10.4 | 69.7 |
| 427 | - | 2 | 5-10 | 41.9 | ... | ... |
| 428 | - | 2 | 5-11 | 72.2 | ... | ... |

| Shectman | Abell | No of levels | min-max $N_{cells}$ | $\theta$ | $\sigma_\theta$ | Med[$\theta$] |
|---|---|---|---|---|---|---|
| 429 | 2366 | 2 | 8     | 133.6 | ... | ... |
| 431 | 2372 | 2 | 5-6   | 32.8  | ... | ... |
| 432 | 2377 | 4 | 14-41 | 122.7 | 14.7 | 127.2 |
| 434 | 2382 | 2 | 7-9   | 33.9  | ... | ... |
| 435 | 2384 | 3 | 6-10  | 132.3 | 39.3 | 126.9 |
| 437 | -    | 1 | 6     | 142.1 | ... | ... |
| 438 | -    | 1 | 5     | 0.1   | ... | ... |
| 439 | 2399 | 3 | 16-23 | 94.4  | 9.2 | 99.6 |
| 440 | 2401 | 4 | 5-18  | 138.8 | 19.6 | 144.0 |
| 441 | 2410 | 3 | 11-22 | 121.9 | 4.1 | 123.9 |
| 443 | 2415 | 2 | 7-11  | 46.8  | ... | ... |
| 444 | -    | 2 | 6-7   | 85.5  | ... | ... |
| 445 | 2420 | 2 | 6-9   | 59.7  | ... | ... |
| 447 | 2426 | 1 | 6     | 23.8  | ... | ... |
| 450 | -    | 1 | 15    | 37.9  | ... | ... |
| 452 | 2448 | 1 | 7     | 59.6  | ... | ... |
| 453 | 2457 | 3 | 7-14  | 25.5  | 3.6 | 25.3 |
| 454 | 2459 | 4 | 7-21  | 144.5 | 5.5 | 145.1 |
| 456 | -    | 1 | 6     | 134.2 | ... | ... |
| 457 | -    | 1 | 11    | 61.1  | ... | ... |
| 461 | -    | 1 | 8     | 34.6  | ... | ... |
| 463 | -    | 3 | 5-20  | 90.2  | 1.3 | 90.0 |
| 464 | 2529 | 1 | 5     | 16.7  | ... | ... |
| 467 | -    | 4 | 8-15  | 52.3  | 17.1 | 43.7 |
| 470 | 2554 | 3 | 5-13  | 110.0 | 4.6 | 112.5 |
| 472 | -    | 2 | 8-10  | 66.8  | ... | ... |
| 473 | -    | 3 | 8-19  | 148.6 | 4.8 | 149.2 |
| 475 | 2569 | 1 | 9     | 59.1  | ... | ... |
| 478 | 2593 | 3 | 10-26 | 179.0 | 4.4 | 178.5 |
| 479 | 2592 | 2 | 6-7   | 74.6  | ... | ... |
| 486 | 2657 | 3 | 5-16  | 84.7  | 14.6 | 82.9 |
| 487 | -    | 3 | 5-11  | 130.3 | 17.5 | 135.6 |
| 488 | -    | 2 | 5-8   | 21.5  | ... | ... |
| 490 | -    | 1 | 6     | 0.1   | ... | ... |

| Shectman | Abell | No of levels | min-max $N_{cells}$ | $\theta$ | $\sigma_\theta$ | Med[$\theta$] |
|---|---|---|---|---|---|---|
| 491 | 2670 | 3 | 8-12 | 33.2 | 4.9 | 32.7 |
| 492 | - | 4 | 8-36 | 39.1 | 12.8 | 37.2 |
| 494 | - | 3 | 7-19 | 178.1 | 18.4 | 3.0 |
| 495 | - | 1 | 8 | 92.9 | ... | ... |
| 496 | 2686 | 3 | 8-22 | 124.9 | 11.2 | 124.6 |
| 498 | - | 3 | 6-12 | 151.8 | 25.1 | 152.9 |
| 500 | 13 | 3 | 7-14 | 83.1 | 28.9 | 89.2 |
| 502 | 16 | 3 | 5-10 | 49.6 | 41.7 | 70.1 |
| 503 | - | 3 | 5-11 | 149.6 | 38.4 | 163.3 |
| 504 | - | 3 | 8-22 | 133.3 | 11.9 | 135.7 |
| 505 | - | 1 | 5 | 70.4 | ... | ... |
| 508 | 23 | 2 | 14-19 | 63.5 | ... | ... |
| 510 | - | 2 | 5-9 | 145.2 | ... | ... |
| 511 | 27 | 4 | 5-40 | 65.2 | 39.2 | 48.6 |
| 512 | - | 4 | 7-34 | 131.0 | 38.7 | 134.9 |
| 513 | 44 | 4 | 6-20 | 153.0 | 18.2 | 147.8 |
| 514 | - | 1 | 8 | 31.6 | ... | ... |
| 517 | - | 2 | 5-11 | 54.1 | ... | ... |
| 518 | - | 3 | 18-44 | 165.8 | 2.6 | 166.1 |
| 520 | - | 2 | 7-14 | 145.6 | ... | ... |
| 522 | - | 2 | 5-10 | 100.2 | ... | ... |
| 524 | 76 | 4 | 5-23 | 17.9 | 14.9 | 20.5 |
| 525 | 84 | 1 | 6 | 133.3 | ... | ... |
| 526 | 85 | 4 | 12-34 | 163.1 | 4.3 | 162.9 |
| 529 | - | 1 | 6 | 42.2 | ... | ... |
| 530 | 93 | 1 | 5 | 161.0 | ... | ... |
| 532 | - | 1 | 12 | 18.2 | ... | ... |
| 533 | 95 | 2 | 10-20 | 45.7 | ... | ... |
| 534 | 98 | 4 | 6-11 | 3.3 | 34.5 | 6.4 |
| 536 | 112 | 1 | 9 | 53.6 | ... | ... |
| 539 | - | 2 | 6-9 | 124.8 | ... | ... |
| 541 | - | 1 | 6 | 46.0 | ... | ... |
| 542 | - | 1 | 6 | 2.6 | ... | ... |
| 543 | 114 | 1 | 6 | 84.6 | ... | ... |

| Shectman | Abell | No of levels | min-max $N_{cells}$ | $\theta$ | $\sigma_\theta$ | Med[$\theta$] |
|----------|-------|--------------|---------------------|----------|-----------------|---------------|
| 544 | - | 1 | 12 | 131.8 | ... | ... |
| 546 | - | 2 | 10-13 | 40.8 | ... | ... |
| 547 | - | 2 | 5-10 | 36.9 | ... | ... |
| 548 | 116 | 3 | 6-28 | 145.9 | 44.0 | 161.2 |
| 549 | 117 | 4 | 15-84 | 35.2 | 17.1 | 32.6 |
| 550 | 119 | 3 | 22-45 | 28.3 | 1.0 | 28.8 |
| 551 | - | 1 | 6 | 12.1 | ... | ... |
| 552 | 120 | 1 | 8 | 66.4 | ... | ... |
| 553 | - | 2 | 6-12 | 15.4 | ... | ... |
| 555 | - | 3 | 5-14 | 97.3 | 10.3 | 95.4 |
| 557 | 126 | 2 | 5-12 | 12.6 | ... | ... |
| 559 | - | 1 | 12 | 15.1 | ... | ... |
| 560 | - | 1 | 5 | 81.9 | ... | ... |
| 561 | - | 1 | 17 | 47.1 | ... | ... |
| 563 | - | 1 | 12 | 150.6 | ... | ... |
| 565 | - | 3 | 7-28 | 94.2 | 34.3 | 83.3 |
| 567 | - | 2 | 7-22 | 87.7 | ... | ... |
| 568 | - | 2 | 7-12 | 77.2 | ... | ... |
| 569 | 147 | 4 | 10-23 | 78.8 | 6.3 | 77.1 |
| 570 | 150 | 3 | 5-10 | 164.2 | 24.1 | 176.7 |
| 571 | 151 | 3 | 29-47 | 40.5 | 13.3 | 47.6 |
| 572 | 154 | 1 | 9 | 172.2 | ... | ... |
| 576 | 160 | 2 | 6-9 | 174.1 | ... | ... |
| 578 | - | 1 | 10 | 0.8 | ... | ... |
| 580 | - | 2 | 7-9 | 22.8 | ... | ... |
| 581 | 168 | 2 | 13-19 | 149.1 | ... | ... |
| 582 | 171 | 1 | 5 | 95.0 | ... | ... |
| 583 | - | 2 | 9-13 | 177.3 | ... | ... |
| 584 | - | 1 | 8 | 35.3 | ... | ... |
| 585 | 175 | 2 | 6-19 | 78.8 | ... | ... |
| 586 | - | 2 | 5-13 | 95.9 | ... | ... |
| 587 | - | 3 | 7-14 | 95.3 | 16.4 | 87.1 |
| 588 | - | 2 | 6-14 | 104.8 | ... | ... |
| 590 | 193 | 3 | 9-14 | 80.1 | 8.4 | 82.5 |

| Shectman | Abell | No of levels | min-max $N_{cells}$ | $\theta$ | $\sigma_\theta$ | Med[$\theta$] |
|---|---|---|---|---|---|---|
| 591 | - | 3 | 9-31 | 59.7 | 5.1 | 56.8 |
| 592 | 194 | 3 | 6-22 | 83.8 | 28.3 | 86.7 |
| 593 | - | 3 | 9-37 | 25.6 | 24.8 | 12.4 |
| 594 | - | 3 | 5-18 | 64.7 | 47.2 | 63.8 |
| 596 | - | 1 | 7 | 68.7 | ... | ... |
| 598 | - | 3 | 5-29 | 67.8 | 22.0 | 56.3 |
| 599 | - | 1 | 13 | 110.4 | ... | ... |
| 600 | - | 1 | 5 | 20.4 | ... | ... |
| 601 | - | 3 | 7-20 | 170.7 | 29.5 | 174.6 |
| 603 | - | 3 | 8-51 | 52.0 | 29.3 | 66.9 |
| 604 | - | 2 | 5-8 | 100.3 | ... | ... |
| 608 | 256 | 1 | 5 | 134.3 | ... | ... |
| 609 | 257 | 3 | 5-17 | 163.2 | 41.4 | 173.5 |
| 612 | 274 | 3 | 6-12 | 73.9 | 2.1 | 72.7 |
| 615 | - | 2 | 8-13 | 89.9 | ... | ... |
| 616 | 295 | 3 | 6-31 | 53.9 | 1.5 | 54.1 |
| 618 | - | 3 | 5-63 | 144.4 | 13.7 | 139.1 |
| 619 | - | 1 | 5 | 74.9 | ... | ... |
| 620 | - | 2 | 6-11 | 46.2 | ... | ... |
| 622 | - | 3 | 6-12 | 35.1 | 5.8 | 38.1 |
| 623 | - | 1 | 6 | 12.4 | ... | ... |
| 624 | - | 1 | 6 | 132.2 | ... | ... |
| 625 | - | 3 | 8-21 | 121.4 | 13.1 | 123.6 |
| 627 | 367 | 3 | 7-11 | 43.8 | 27.7 | 44.8 |
| 628 | - | 2 | 5-7 | 59.1 | ... | ... |
| 629 | - | 2 | 11-17 | 139.6 | ... | ... |
| 630 | - | 3 | 5-14 | 12.6 | 30.4 | 12.7 |
| 631 | - | 1 | 6 | 135.4 | ... | ... |
| 632 | - | 2 | 6-10 | 67.0 | ... | ... |
| 633 | 400 | 4 | 5-42 | 90.1 | 13.1 | 96.1 |
| 635 | - | 1 | 10 | 56.8 | ... | ... |
| 638 | - | 3 | 6-21 | 1.7 | 2.3 | 0.5 |
| 639 | 415 | 3 | 6-9 | 48.7 | 15.7 | 43.7 |
| 640 | - | 3 | 10-27 | 95.3 | 10.8 | 96.8 |

| Shectman | Abell | No of levels | min-max $N_{cells}$ | $\theta$ | $\sigma_\theta$ | Med$[\theta]$ |
|---|---|---|---|---|---|---|
| 641 | 420 | 1 | 6 | 134.6 | ... | ... |
| 642 | 423 | 1 | 6 | 91.7 | ... | ... |
| 646 | - | 2 | 6-8 | 43.3 | ... | ... |
| 647 | - | 2 | 5-11 | 74.1 | ... | ... |
| 648 | - | 1 | 8 | 159.1 | ... | ... |
| 649 | - | 2 | 5-10 | 131.8 | ... | ... |

**Table 1b**: Position angles of Abell clusters, identified at the 3.6 level but estimated at the 2.5 level.

| Abell | $N_{cells}$ | $\theta$ | Abell | $N_{cells}$ | $\theta$ | Abell | $N_{cells}$ | $\theta$ |
|-------|-------------|----------|-------|-------------|----------|-------|-------------|----------|
| 779   | 5  | 72.2  | 1518 | 5  | 69.9  | 148  | 8  | 56.2  |
| 1069  | 8  | 138.7 | 1524 | 10 | 8.8   | 152  | 10 | 110.8 |
| 1170  | 5  | 158.7 | 1729 | 14 | 50.5  | 225  | 5  | 72.2  |
| 1174  | 6  | 147.5 | 1796 | 5  | 109.1 | 403  | 5  | 69.6  |
| 1189  | 6  | 167.3 | 1818 | 10 | 107.5 | 2400 | 9  | 142.9 |
| 1227  | 9  | 145.0 | 1921 | 8  | 179.9 | 2412 | 5  | 89.7  |
| 1278  | 5  | 118.9 | 1926 | 6  | 47.9  | 2421 | 6  | 1.3   |
| 1344  | 10 | 127.8 | 1953 | 14 | 111.0 | 2490 | 6  | 18.2  |
| 1411  | 5  | 44.0  | 1982 | 13 | 59.9  | 2525 | 5  | 17.0  |
| 1427  | 5  | 160.4 | 2120 | 5  | 18.1  | 2665 | 5  | 108.6 |
| 1515  | 15 | 161.8 | 2153 | 6  | 119.6 |      |    |       |

**Table 1c**: Position angles of Abell clusters, identified at the 3.6 level but estimated at the 1.8 level.

| Abell | $N_{cells}$ | $\theta$ | Abell | $N_{cells}$ | $\theta$ | Abell | $N_{cells}$ | $\theta$ |
|-------|-------------|----------|-------|-------------|----------|-------|-------------|----------|
| 865   | 6  | 12.5  | 1741 | 5  | 109.3 | 2208 | 5  | 20.0  |
| 903   | 33 | 117.5 | 1768 | 7  | 133.6 | 65   | 10 | 87.0  |
| 985   | 11 | 105.5 | 1845 | 7  | 35.5  | 94   | 6  | 71.4  |
| 987   | 7  | 171.6 | 1861 | 5  | 18.7  | 144  | 5  | 18.0  |
| 1092  | 5  | 109.0 | 1870 | 10 | 72.1  | 211  | 8  | 163.6 |
| 1201  | 23 | 166.8 | 1936 | 6  | 18.9  | 212  | 5  | 84.7  |
| 1242  | 6  | 17.9  | 1956 | 6  | 133.6 | 240  | 9  | 36.0  |
| 1262  | 5  | 71.1  | 1990 | 5  | 70.2  | 243  | 5  | 160.8 |
| 1292  | 7  | 177.4 | 2021 | 6  | 13.1  | 2403 | 11 | 36.4  |
| 1327  | 5  | 146.3 | 2025 | 7  | 83.1  | 2477 | 6  | 91.5  |
| 1387  | 8  | 51.7  | 2100 | 6  | 14.2  | 2638 | 8  | 32.0  |
| 1441  | 9  | 153.0 | 2106 | 6  | 93.6  | 2676 | 8  | 157.7 |
| 1625  | 8  | 150.2 | 2195 | 13 | 85.6  | 2703 | 5  | 71.2  |
| 1688  | 5  | 89.7  | 2196 | 5  | 70.6  |      |    |       |

**Table 1d**: Position angles of Abell clusters, identified at the 2.5 level but estimated at the 1.8 level. Note that there are 12 clusters in common with table 1b, which for comparison reasons we have not averaged their estimated position angles.

| Abell | $N_{cells}$ | $\theta$ | Abell | $N_{cells}$ | $\theta$ | Abell | $N_{cells}$ | $\theta$ |
|---|---|---|---|---|---|---|---|---|
| 716  | 5   | 69.8  | 1444 | 5  | 71.6  | 2148 | 8  | 90.5  |
| 733  | 6   | 91.0  | 1466 | 8  | 115.2 | 2169 | 16 | 50.2  |
| 779  | 13  | 77.6  | 1474 | 9  | 67.2  | 2175 | 9  | 103.9 |
| 795  | 6   | 129.4 | 1495 | 21 | 144.9 | 2177 | 13 | 53.2  |
| 803  | 8   | 47.7  | 1518 | 10 | 90.8  | 48   | 6  | 130.0 |
| 866  | 7   | 82.3  | 1526 | 8  | 138.7 | 53   | 10 | 119.6 |
| 878  | 17  | 97.1  | 1534 | 9  | 12.6  | 103  | 10 | 72.3  |
| 929  | 9   | 84.8  | 1573 | 9  | 119.9 | 111  | 10 | 3.3   |
| 967  | 6   | 109.1 | 1581 | 5  | 163.1 | 13   | 6  | 145.3 |
| 992  | 370 | 170.8 | 1583 | 5  | 162.4 | 172  | 15 | 87.0  |
| 1003 | 17  | 119.8 | 1595 | 14 | 22.3  | 179  | 12 | 136.2 |
| 1022 | 11  | 58.4  | 1599 | 7  | 177.2 | 195  | 5  | 159.9 |
| 1051 | 6   | 90.4  | 1606 | 10 | 78.7  | 207  | 8  | 12.1  |
| 1069 | 12  | 179.1 | 1616 | 8  | 131.5 | 225  | 21 | 41.5  |
| 1097 | 11  | 86.1  | 1630 | 5  | 160.1 | 229  | 8  | 22.4  |
| 1098 | 16  | 121.3 | 1684 | 5  | 19.2  | 267  | 26 | 94.7  |
| 1108 | 10  | 48.1  | 1690 | 9  | 101.2 | 292  | 7  | 84.5  |
| 1109 | 30  | 7.1   | 1693 | 11 | 155.7 | 358  | 10 | 9.7   |
| 1118 | 6   | 40.5  | 1696 | 11 | 6.3   | 395  | 16 | 19.6  |
| 1132 | 10  | 110.2 | 1715 | 5  | 162.2 | 403  | 12 | 59.5  |
| 1135 | 14  | 44.9  | 1729 | 31 | 47.9  | 410  | 6  | 91.9  |
| 1141 | 6   | 88.1  | 1769 | 15 | 75.3  | 428  | 11 | 123.4 |
| 1143 | 42  | 13.4  | 1808 | 6  | 88.4  | 2346 | 7  | 45.6  |
| 1152 | 7   | 58.7  | 1823 | 6  | 42.8  | 2361 | 6  | 87.1  |
| 1170 | 11  | 162.2 | 1844 | 14 | 12.7  | 2365 | 6  | 46.2  |
| 1179 | 11  | 49.1  | 1850 | 18 | 54.4  | 2412 | 13 | 117.0 |
| 1189 | 32  | 10.9  | 1860 | 8  | 134.7 | 2490 | 13 | 16.0  |
| 1198 | 11  | 164.3 | 1891 | 6  | 133.5 | 2495 | 6  | 42.5  |
| 1216 | 7   | 179.4 | 1909 | 7  | 80.8  | 2511 | 7  | 136.9 |
| 1218 | 9   | 175.4 | 1944 | 13 | 42.7  | 2525 | 12 | 144.5 |
| 1257 | 12  | 25.0  | 1960 | 16 | 108.9 | 2528 | 10 | 116.8 |
| 1264 | 5   | 74.7  | 1976 | 7  | 46.8  | 2549 | 20 | 165.6 |
| 1375 | 5   | 0.7   | 1999 | 9  | 21.0  | 2583 | 13 | 11.7  |
| 1390 | 5   | 95.4  | 2019 | 8  | 170.8 | 2589 | 11 | 14.5  |

| Abell | $N_{cells}$ | $\theta$ | Abell | $N_{cells}$ | $\theta$ | Abell | $N_{cells}$ | $\theta$ |
|---|---|---|---|---|---|---|---|---|
| 1408 | 6 | 89.7 | 2026 | 6 | 46.3 | 2614 | 9 | 84.4 |
| 1409 | 9 | 36.2 | 2030 | 6 | 179.5 | 2630 | 7 | 167.2 |
| 1412 | 5 | 20.7 | 2034 | 11 | 48.8 | 2654 | 10 | 24.1 |
| 1419 | 14 | 33.2 | 2064 | 6 | 134.5 | 2656 | 14 | 26.3 |
| 1423 | 11 | 130.4 | 2101 | 6 | 91.4 | 2665 | 15 | 122.1 |
| 1433 | 8 | 5.6 | 2120 | 14 | 174.0 | 2678 | 5 | 70.5 |
| 1437 | 5 | 58.2 | 2141 | 8 | 89.2 | 2698 | 9 | 19.1 |

**Table 2:** Nearest-neighbour alignment signal and its standard deviation (column 2) as a function of limiting maximum cluster separation (column 1). Column 3 shows the signal to noise ratio, column 4 the number of separations considered and column 5 the probability, estimated from a Kolmogorov-Smirnov test, that the distribution of $\phi_{nn}$'s is drawn from a uniform one.

| $D_{lim}$ ($h^{-1}$ Mpc) | $\delta \pm \sigma$ | $\delta/\sigma$ | No of pairs | $P_{KS}$ |
|---|---|---|---|---|
| 10 | -9.4 ± 3.3 | 2.9 | 61 | 0.04 |
| 15 | -6.3 ± 2.5 | 2.5 | 106 | 0.03 |
| 30 | -3.9 ± 1.9 | 2.0 | 192 | 0.22 |
| 50 | -4.3 ± 1.6 | 2.7 | 254 | 0.06 |
| 60 | -4.4 ± 1.6 | 2.8 | 260 | 0.05 |

**Table 3:** All-neighbour alignment signal.

| $D_{lim}$ ($h^{-1}$ Mpc) | $\delta \pm \sigma$ | $\delta/\sigma$ | No of pairs | $P_{KS}$ |
|---|---|---|---|---|
| 0 - 10 | -10.9 ± 2.9 | 3.8 | 80 | $2 \times 10^{-3}$ |
| 10 - 30 | -2.45 ± 1.13 | 2.2 | 528 | 0.06 |
| 30 - 50 | -2 ± 0.85 | 2.3 | 904 | 0.015 |
| 50 - 150 | -1.5 ± 0.41 | 3.6 | 3944 | 0.01 |
| 0 - 150 | -1.8 ± 0.35 | 5.1 | 5456 | $10^{-4}$ |

# Figure Captions

**Figure 1.** Cluster position angles, $\theta$, as a function of the number of $10' \times 10'$ cells, defining the cluster sampling area, used in their estimation. The position angles estimated only at one overdensity level, are plotted.

**Figure 2.** As in Figure 1 but for clusters with more than one position angle determination: **(a.)** The average position angle, $\langle \theta \rangle$ as a function of minimum number of $10' \times 10'$ cells used in its estimation. **(b.)** $\langle \theta \rangle$ as a function of the maximum number of cells used.

**Figure 3.** The frequency distribution of $\phi_{nn}$: **(a.)** Solid and broken lines corresponds to nearest-neighbour separations of $D \leq 15$ and $\leq 10 \ h^{-1}$ Mpc, respectively. **(b.)** $\phi_{nn}$ distribution for all nearest-neighbour separations.

**Figure 4.** The frequency distribution of all-neighbour relative angles, $\phi_{ij}$, at the four indicated separation ranges.